\documentclass[conference]{IEEEtran}
\IEEEoverridecommandlockouts
\usepackage{cite}
\usepackage{amsmath,amssymb,amsfonts}
\usepackage{algorithmic}
\usepackage{graphicx}
\usepackage{textcomp}
\usepackage{xcolor}
\usepackage[hyphens]{url}
\usepackage{hyperref}
\usepackage{booktabs}
\usepackage{mdframed}
\usepackage[]{footmisc}
\usepackage{footnote}
\usepackage[a4paper, total={184mm,239mm}]{geometry}
\def\BibTeX{{\rm B\kern-.05em{\sc i\kern-.025em b}\kern-.08em
    T\kern-.1667em\lower.7ex\hbox{E}\kern-.125emX}}
\usepackage{algorithm2e}
\hypersetup{
    colorlinks=true,
    linkcolor=blue,
    filecolor=magenta,      
    urlcolor=cyan,
    citecolor = magenta,
    pdfpagemode=FullScreen,
}

\newcommand*\design{{FeReX}\xspace}
\newcommand*\sch{{sch}\xspace}
\newcommand*\sto{{sto}\xspace}
\PassOptionsToPackage{bookmarks={false}}{hyperref}

\renewcommand{\IEEEbibitemsep}{0pt plus 0.5pt}
\makeatletter
\IEEEtriggercmd{\reset@font\normalfont\fontsize{7.0pt}{7.0   pt}\selectfont}
\makeatother
\IEEEtriggeratref{1}

\begin{document}
\bstctlcite{IEEEexample:BSTcontrol}
\title{FeReX: A Reconfigurable Design of Multi-bit Ferroelectric Compute-in-Memory for Nearest Neighbor Search}
 \author{
 \small
 Zhicheng Xu$^{1,2}$, Che-Kai Liu$^3$, Chao Li$^2$, Ruibin Mao$^1$, Jianyi Yang$^{2,*}$,\\
 Thomas K{\"a}mpfe$^4$, Mohsen Imani$^5$, Can Li$^{1*}$, Cheng Zhuo$^{2,6,*}$ and Xunzhao Yin$^{2,6,*}$\\
$^1$Department of Electrical and Electronic Engineering, The University of Hong Kong, Hong Kong\\
$^2$Zhejiang University, Hangzhou,  China\\
$^3$School of Electrical and Computer Engineering, Georgia Institute of Technology, USA\\
$^4$ Center Nanoelectric Technologies, Fraunhofer IPMS, Dresden, Germany\\ 
$^5$Department of Computer Science, University of California Irvine, USA\\
$^6$Key Laboratory of Collaborative Sensing and Autonomous Unmanned Systems of Zhejiang Province, Hangzhou, China\\
$^*$Corresponding authors, email: canl@hku.hk; \{yangjy, czhuo, xzyin1\}@zju.edu.cn
 \vspace{-4ex}
}

\maketitle
\begin{abstract} 

Rapid advancements in artificial intelligence have given rise to transformative models, profoundly impacting our lives. These models demand massive volumes
of data to operate effectively, exacerbating the data-transfer bottleneck inherent in the conventional von-Neumann architecture. Compute-in-memory (CIM), a novel computing paradigm, tackles these issues by seamlessly embedding in-memory search functions, thereby obviating the need for data transfers. However, existing non-volatile memory (NVM)-based accelerators are application specific. During the similarity based associative search operation, they only support a single, specific distance metric, such as Hamming, Manhattan, or Euclidean distance in measuring the query against the stored data, calling for 
reconfigurable in-memory solutions adaptable to various applications. 
To overcome such a limitation, in this paper, we present FeReX, a reconfigurable associative memory (AM) that accommodates various distance metrics including Hamming, Manhattan, and Euclidean distances. Leveraging multi-bit ferroelectric field-effect transistors (FeFETs) as the proxy and a hardware-software co-design approach, we introduce a constrained satisfaction problem (CSP)-based method to automate AM search input voltage and stored voltage configurations for different distance based search functions. Device-circuit co-simulations first validate the effectiveness of the proposed \design methodology for reconfigurable search distance functions. Then, we benchmark FeReX in the context of k-nearest neighbor (KNN) and hyperdimensional computing (HDC), which highlights the robustness of \design and demonstrates up to 250$\times$ speedup and $10^4$ energy savings compared with GPU.

\end{abstract}


\section{Introduction}
\label{sec:intro}
The  artificial intelligence (AI) 
models  yield a profound influence over various aspects of our lives. These models, however, frequently require vast amounts of data for their operation, thus exacerbating the data-transfer bottleneck inherent in the traditional von Neumann architecture.
Consequently, 
there is a growing demand for a departure from the conventional computing paradigm, one that seamlessly integrates the critical functionalities of emerging AI models within the memory itself. This shift is not only desirable but also essential to keep pace with the demands of modern computing.

Compute-in-memory (CIM) has emerged as an alternative computing paradigm that integrates the separated computing unit and memory that exists in Von Neuman machine altogether \cite{li2022imars, wei2023imga, yan2022swim, chen2022accelerating, yan2022computing}. 
Several CIM primitives, i.e., associative memories (AMs) that support various distance metric computations between input and stored vectors have demonstrated their potential for accelerating similarity based inferences  in novel machine learning algorithms \cite{hu2021memory, lele2023heterogeneous, yin2022ferroelectric,  peng2020dnn+, cai2022energy, wang2021triangle, yin2020fecam}. 
Hamming distance (HD)-based CIM design has been originally proposed \cite{ni2019ferroelectric} for memory-augmented neural networks (MANN), but it suffers from non-negligible classification accuracy degradation. Recently, CIM design that implements Manhattan distance for MANN classification has been experimentally verified \cite{li2021sapiens}, and CIM design realizing Euclidean distance for hyperdimensional computing (HDC) has been demonstrated at the device level \cite{kazemi2022achieving}. These CIM based AM designs aim to address the non-negligible algorithmic accuracy degradation with complex distance functions used in a certain application.
However, existing non-volatile memory (NVM)-based AMs are limited to a specific classification task, as one AM design can only support a single distance computation, such as Hamming \cite{ni2019ferroelectric, liu2023reconfigurable, xu2023challenges}, Manhattan \cite{li2021sapiens}, Euclidean \cite{kazemi2022achieving}, and sigmoid \cite{kazemi2021fefet}.
A CIM search engine that can achieve a reconfigurable distance function is highly desirable. 
Based on the nature of various applications, different distance functions may be used during the similarity based search, and, within a certain application, several distance functions may be exploited for various datasets.

In this paper, we propose FeReX, a reconfigurable CIM-based AM for Hamming, Manhattan, and Euclidean distance searches, utilizing multi-bit ferroelectric field-effect transistor (FeFET) devices   as the proxy. 
We propose a hardware-software co-design scheme  to efficiently realize similarity searches  between a query and stored vectors in terms of various distance metrics.
This involves constructing a matrix of target distance values  between the query and stored vectors based on the given distance function. 
To accommodate this target matrix, 
we formulate 
a constrained satisfaction problem (CSP), which incorporates   the FeFET device and crossbar constraints  related to the output currents, input voltages and stored threshold voltages.
By solving the CSP using backtracking and AC-3 algorithm, we find the optimal
search input  and stored voltage configurations for the input query and stored vectors that align the CSP  formulation with the target distance matrix.
In this sense, FeReX can be readily configured to support a range of distance functions in an automated way. 
FeReX incorporates a Loser-Take-All (LTA) circuit structure, enabling it to support nearest neighbor search functionality. Our extensive performance assessment in the realms of k-nearest neighbor (KNN) and HDC applications underscores the robustness and efficacy of our design approach, highlighting its resilience and efficiency. Notably, FeReX achieves up to 250× speedup and $10^4$ energy savings compared to GPU implementations. 
To  best of our knowledge, this work represents the first  reconfigurable distance search implementation within NVM-based AM.
\section{Background}
\label{sec:background}
In this Sec., we review the FeFET  characteristics that are exploited within the \design. Then, recent AM designs for NN search are briefly summarized.

\subsection{FeFET Characteristics}
\begin{figure}[!t]
\centering
\includegraphics[width=\linewidth]{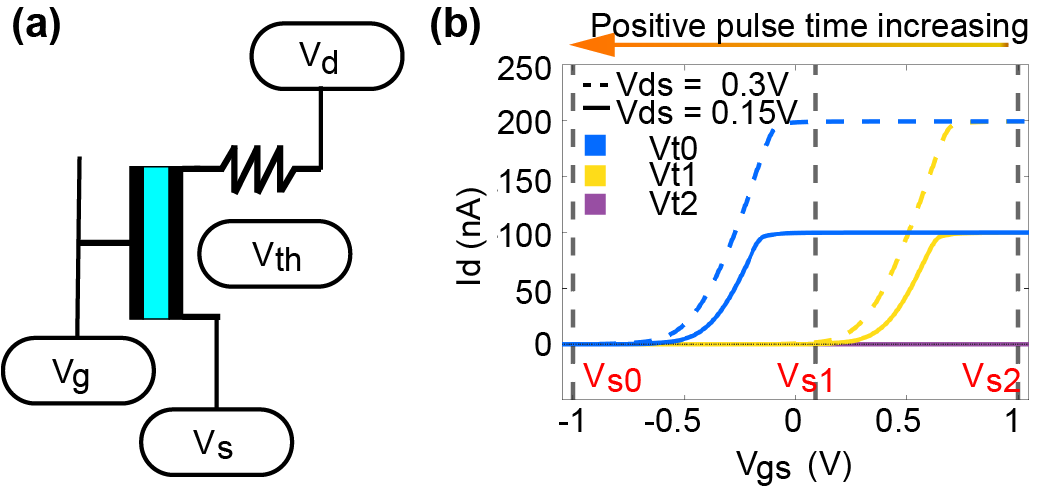}
\caption{{\bf (a)} 1FeFET1R structure. {\bf (b)} multi-level I-V curve of 1FeFET1R, where $V_{t0}$, $V_{t1}$, $V_{t2}$ represent different $V_{th}$ stored in the FeFET,  $V_{s0}$, $V_{s1}$, $V_{s2}$ represent different search voltage (i.e, $V_{gs}$) applied to the FeFET, and two different $V_{ds}$ result in two level of ON currents.}
\vspace{-2ex}
\label{fig:FeFETBasic}
\end{figure}
Excellent CMOS compatibility, outstanding scalability, and superior energy efficiency \cite{boscke2011ferroelectricity} of HfO$_\text{2}$ ferroelectric materials elucidate the competitiveness of Ferroelectric FET (FeFET) among other NVMs. Based on the conventional CMOS transistor, a FeFET is made with ferroelectric materials integrated into the CMOS gate stack. The stored value is represented by the threshold voltage ($V_{th}$) of a FeFET, and can be altered by applying a positive or negative voltage pulse at the device gate, which in turn changes the polarization of the Fe layer. Specifically, the value of $V_{th}$ is determined by the duration and magnitude of the applied voltage pulse \cite{ni2019ferroelectric}. For instance, if the duration of a given positive voltage pulse increases, the $V_{th}$ will shift lower accordingly.

Recently, Soliman et al. propose a cell that integrates a resistor with a single FeFET \cite{soliman2020ultra}, as shown in Fig~\ref{fig:FeFETBasic}(a). It is demonstrated in \cite{soliman2020ultra,yin2023ultracompact} that by connecting a large resistor at the source (or equivalently, drain)  of the FeFET, the ON state current $I_{ds}$ is significantly reduced and thus is independent of  $V_{th}$ variation \cite{soliman2020ultra}. 
Saito et al. further demonstrate a back-end-of-line (BEOL) 1FeFET1R structure, incurring no additional area penalty with an $M\Omega$ resistor integrated with a FeFET \cite{saito2021analog}. Given a $V_{ds}$ and resistance $R$,  
The conducting current of a FeFET can be approximated as Min\{$I_{sat}$,$V_{ds}/R$\} due to the fact that it is possible when $I_{ds} = V_{ds}/R$ under a given $V_{gs}$, the FeFET operates in the linear region. In this work, all $V_{ds}$ values are integer multiples of the minimum $V_{ds}$ value, ensuring that all $I_{ds}$ values are interger multiples of the minimum $I_{ds}$ value. 
Fig.~\ref{fig:FeFETBasic}(b) illustrates a multi-level cell (MLC) 1FeFET1R characteristics. When $V_{gs} > V_{th}$, various $V_{th}$ and $V_{gs}$ values can be explored, where $I_{ds}$ is approximately equivalent to $V_{ds}/R$, while $I_{ds}$ approaches 0 if $V_{gs} < V_{th}$.
\subsection{Existing AM Designs}
\begin{table}[]
\centering
\caption{Existing AMs with Different Distance Functions}
\label{tab:existing}
\resizebox{1\columnwidth}{!}{
\begin{tabular}{|c|cccc|}
\toprule
\textbf{Design}& NVM & Cell structure & MLC & Distance function \\ \midrule
\textbf{Nat. Ele. \cite{karunaratne2020memory}} & PCM& 1PCM&No & Hamming \\
\textbf{IEDM'20 \cite{li2020scalable}} & FeFET & 2FeFET-1T& Yes &Best-match   \\
\textbf{TED'21 \cite{li2021sapiens}} & RRAM& 2RRAM& Yes& Manhattan\\
\textbf{TC'21 \cite{kazemi2021fefet}} & FeFET & 2FeFET & Yes &Sigmoid\\ 
\textbf{SR'22 \cite{kazemi2022achieving}} & FeFET &2FeFET& Yes &Euclidean\\
\midrule
\textbf{FeReX (This work)} & FeFET & 1FeFET-1R & Yes & HD/$L_1$/$L_2$\\
\bottomrule
\end{tabular}
}
\end{table}
AM has been deployed in a variety of scenarios such as HDC \cite{karunaratne2020memory,shou2023see, huang2023fefet}, MANN \cite{li2021sapiens}, few-shot learning \cite{kazemi2021fefet}, and so on. Table~\ref{tab:existing} summarizes existing AMs based on single-level cell/multi-level cell (SLC/MLC) NVMs with different distance functions. A matching-based MLC 2FeFET-1T AM has been fabricated in \cite{li2020scalable}. To further achieve algorithmic level accuracy, AMs with intricate distance functions utilizing MLC cells have been proposed including sigmoid and Euclidean functions \cite{kazemi2021fefet,kazemi2022achieving}, etc. However, these efforts are typically designed for a fixed distance function. In this work, FeReX is able to support multiple distance functions as shown in Tab.~\ref{tab:existing}. Below we elaborate on the designed AM and its peripherals first. Then, the proposed encoding scheme  for selecting the search input voltages and programming the $V_{th}$ voltages is elucidated in Sec. \ref{sec:algorithm}.
\section{FeReX: Reconfigurable In-Memory Search Engine}
\label{sec:cim}

\begin{figure}
    \centering
    \includegraphics[width=\linewidth]{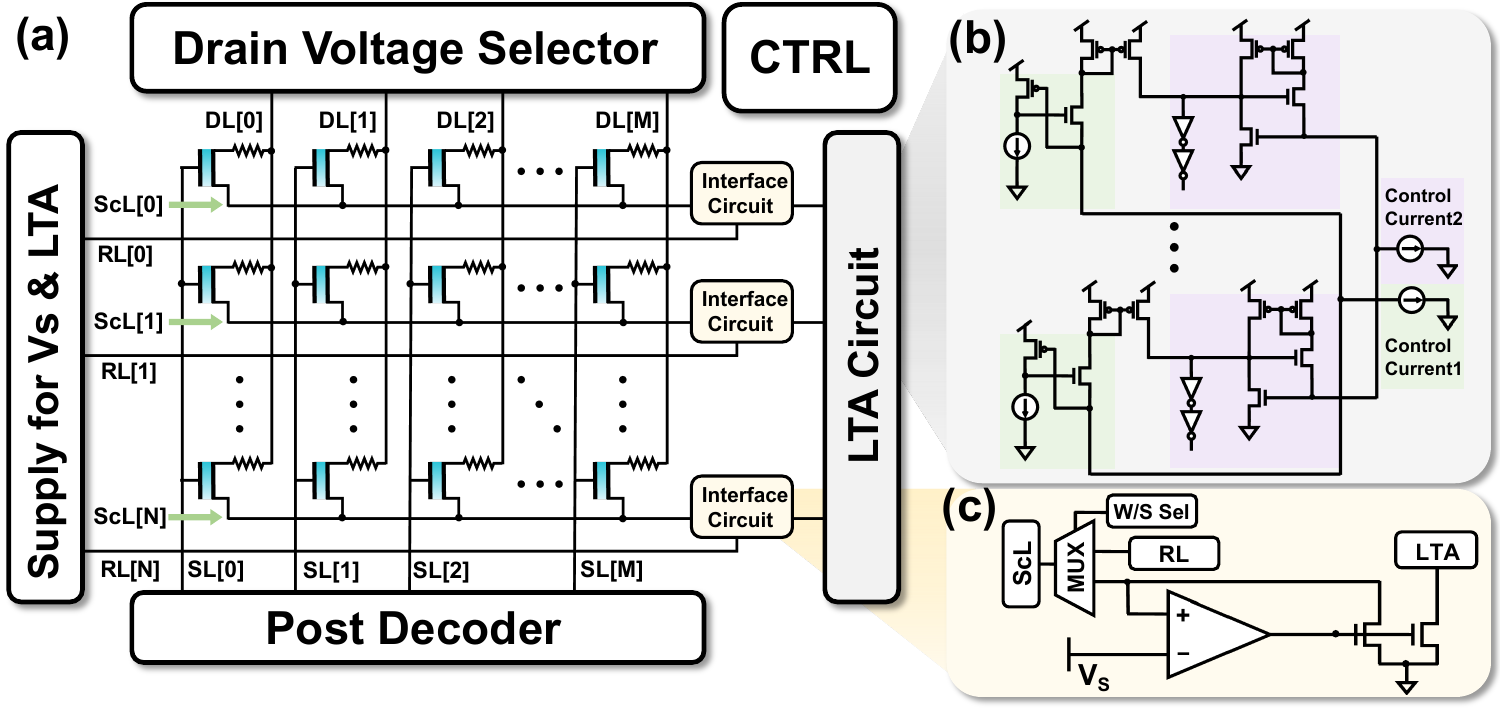}
    \vspace{-2ex}
    \caption{(a) FeReX AM overview.  (b) LTA and (c) Interface circuit.}
    \label{fig:overview}
    \vspace{-2ex}
\end{figure}
\subsection{FeReX Circuit Design}

In this subsection, we briefly describe FeReX, the  FeFET-based AM design along with its distance sensing peripherals. 
The  peripherals for the  array 
includes the level shifters for high write voltages, column switch matrix for selecting columns and input decoder (or digital-to-analog converter) \cite{chen2018neurosim}.
Fig.~\ref{fig:overview} shows the detailed circuit schematic of the proposed FeReX. FeReX consists of a 1FeFET1R based crossbar array with the drain voltage selector and the interface sensing circuit blocks for each row. 
The loser-take-all (LTA) circuitry compares the currents from array rows to perform nearest neighbor search operation.  The search lines (SLs) and drain lines (DLs) are shared by the FeFETs within the same column, and the source lines (ScLs) link the FeFETs within the same row, as shown in Fig.~\ref{fig:overview}(a). 

During the write/erase phase, the MUX of interface circuit selects row lines (RLs),  and  $V_{ScL}=V_{RL}$. In this configuration, the RL voltage of the selected row is 0V, while the RL voltage of the unselected rows  is raised to half of  $V_{write}/V_{erase}$. 
Such writing inhibition scheme prevents write disturbance\cite{ni2018write}.
During the search phase, 
search voltages are applied to   FeFET gates through SLs, and the MUX in the interface circuit selects the op-amp, setting all voltages on ScLs  to $V_s$.
As can be seen from Fig. \ref{fig:FeFETBasic},
the FeFET's ON state current flows from the DL to the ScL only when the applied search voltage $V_{search}$ at the gate exceeds the stored  threshold voltage $V_{th}$. 
Otherwise, the FeFET remains in the cut-off state. 
The ON  current $I_{ON}$ through the FeFET is determined by the voltage drain-source voltage $V_{ds}$, 
as discussed in Sec. \ref{sec:background}.
Given that all FeFETs sharing the same DL experience the same $V_{ds}$,  the ON currents $I_{ON}$ through the activated FeFETs within the same column are identical. 
The currents flowing through FeFETs in the same row are aggregated at ScL and sensed by interface circuit.
The op-amps of all rows are used to 
inhibit ScL voltage fluctuation, 
as the change in $V_{ds}$ of FeFETs will alter the $I_{ON}$ accordingly, resulting in inaccurate LTA sensing.
LTA circuit compares the row currents and indicates the rwo with the minimal current.
The operation of current domain LTA circuitry is similar to winner-take-all (WTA), which has been utilized for NN detection as well. 
Readers interested in detailed explanations can refer to \cite{liu2022cosime}.


\subsection{FeReX Encoding Algorithm}
\label{sec:algorithm}
\begin{figure}
    \centering
    \includegraphics[width=\linewidth]{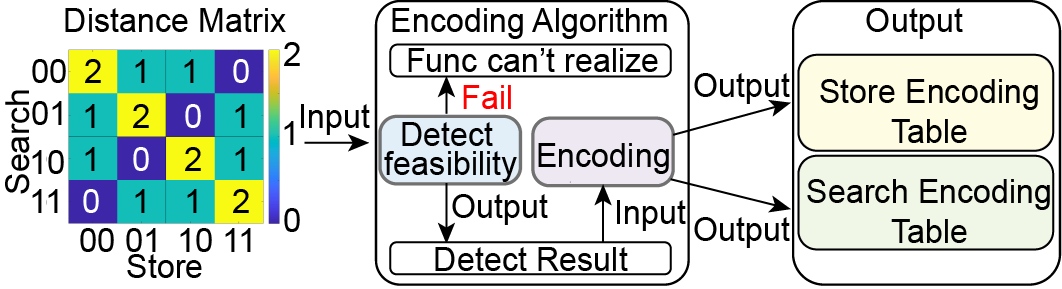}
    \caption{Workflow of  FeReX's encoding scheme.}
    \label{fig:flow}
\end{figure}

Fig.~\ref{fig:flow} depicts the overview of our proposed encoding algorithm that finds the search and stored
voltage configurations for  FeReX  array given a distance function.
The target distance values between the query and stored vectors are first constructed as a function table matrix.
Then a CSP incorporating the FeFET  constraints is formulated to determine whether the target distance matrix can be achieved using  FeReX array. 
The  query  and stored voltage configurations are  encoded by addressing the CSP to align with the target distance matrix.

\begin{figure*}[!t]
  \centering
  
  \includegraphics[width=1.95\columnwidth]{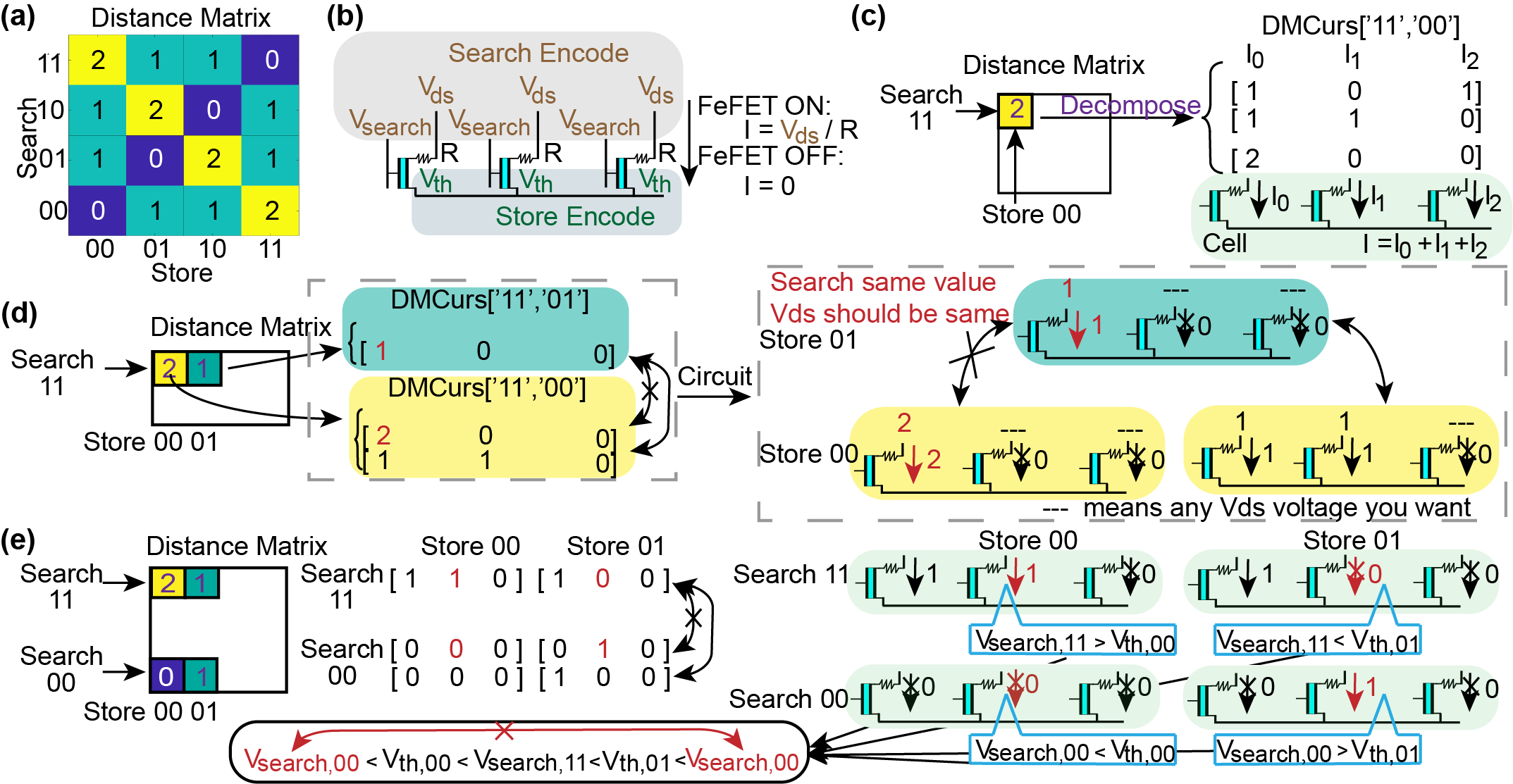}
  
  \caption{(a) DM of 2-bit Hamming Distance. (b) Encoding with FeReX circuit. The stored encoding corresponds to  programmed $V_{th}$ values, while the search encoding corresponds to  FeFET's $V_{ds}$ and $V_{gs}$ voltages. (c) DM element decomposition process
  based on the number of FeFETs  in an AM cell. (d) and (e) The two constraint examples, where (d)  for the same search voltage, the current of an FeFET must either be  identical or 0, and (e) if $\text{FeFET}_{Search11,Store00,2}$ is ON, $\text{FeFET}_{Search11,Store01,2}$ is OFF, a conflict occurs if $\text{FeFET}_{Search00,Store00,2}$ is OFF and $\text{FeFET}_{Search00,Store01,2}$ is ON.}
  \label{fig:constraints}
    \vspace{-2ex}
\end{figure*}
\label{sec:Encoding Algorithm}
Unlike conventional AM designs that consist of a fixed number of NVM devices per cell, \design  flexibly  configures the number of FeFETs  in each AM cell to represent the data vectors. 
The distance metrics  can be represented by the Distance Matrix (DM). 
Within the matrix, columns stand for  stored values, and rows correspond to various search values, with each element in the matrix denoting the distance between a stored value and a search value. Fig.~\ref{fig:constraints}(a) shows the DM  corresponding to a 2-bit Hamming distance, i.e., the distance between the input search vector `00' and  store vector `11' is 2.







Fig.~\ref{fig:constraints}(b) illustrates the search and stored data  encoding to the FeReX circuit. 
The stored encoding is represented by $V_{th}$ value in each FeFET device, while the search encoding consists of the FeFET's $V_{ds}$ and $V_{gs}$ voltages. 
$V_{ds}$   determines the current flowing through the FeFET when the FeFET is activated as shown in Fig. \ref{fig:FeFETBasic}(b). 
The total current flowing through a cell of FeReX represents the distance value between the stored value and input value, i.e., the DM element value.
In order to implement the DM using the FeReX cell, we need to figure out the search and stored encoding configuration within a cell.
Without loss of generality, the number of FeFETs per cell is $k$, the DM element value at  row $\sch$ and column $\sto$ is denoted as $I_{\sch,\sto}$, and the current flowing through the FeFET $i$ under  search $\sch$ and stored $\sto$ value condition  is $I_{\sch,\sto,i}$.
Implementing the DM based on a FeReX cell involves solving a constrained satisfaction problem (CSP) with three  constraints. 

Fig.~\ref{fig:constraints}(c) illustrates the representation of  a DM element $I_{\sch,\sto}$ `2'  by the currents of a set of FeFETs DMCurs[$\sch$, $\sto$] (three FeFETs are used in this example). The implementation decomposes the element $I_{\sch,\sto}$ into decomposed values, i.e.,  $I_{\sch,\sto}=\sum_{i=1}^{k}I_{\sch,\sto,i}$, where 
$I_{\sch,\sto,i}$=`0' indicates  FeFET $i$ is at OFF-state, and $I_{\sch,\sto,i}$=`1/2' indicates that FeFET $i$ is activated with multi-level $V_{ds}$s as shown in Fig. \ref{fig:FeFETBasic}(b).
Since the current of FeFET $i$ under stored $\sto$ and search $\sch$ condition, i.e.,  $I_{\sch,\sto,i}$ represents the value between '0' and the maximal DM value, and the number of possible $I_{\sch,\sto,i}$ values is limited per FeFET's operating condition, the possible FeFET currents $I_{\sch,\sto,i}$  representing DM element $I_{\sch,\sto}$ are constrained, forming the set DMCurs[$\sch$, $\sto$].
We refer to this constraint as the first constraint.


Secondly, considering that the  FeFET $i$ under  search $\sch$ condition should either conduct the identical ON current, or be at  OFF state, the current of this FeFET under different $\sto$ conditions should be the same or 0, i.e., $I_{\sch,\sto_a,i}=I_{\sch,\sto_b,i}$ or 0, $\forall{a,b \in sto}$.
For example, as shown in Fig.~\ref{fig:constraints}(d), $I_{Search11,Store00,1}$ must be equal to $I_{Search11,Store01,1}$ or 0. We refer to this as the second constraint.

The third constraint arises from the multi-level nature of FeFETs and can be expressed as follows: 
The FeFET only turns ON when its $V_{gs}>V_{th}$, therefore, if applying the same search $\sch$ but different store $\sto$ conditions results in different conducting states,i.e., $I_{\sch,\sto_a,i} \neq 0$ and $I_{\sch,\sto_b,i} = 0$, then the voltages corresponding to the search and store conditions must satisfy: $V_{\sto_a}<V_{\sch}<V_{\sto_b}$.
This stored threshold voltage relation must be satisfied when applying different search $\sch'$ condition, i.e., $I_{\sch',\sto_a,i} \geq I_{\sch',\sto_b,i}$.
For example, the DMcurs values for the $2\ $FeFET under $Search11Store00$, $Search11Store01$, $Search00Store00$, $Search00Store01$
as shown in Fig.~\ref{fig:constraints}(e) results in a conflict, i.e., $V_{search,00} < V_{search,00} $.
We refer to this as the third constraint.
\begin{algorithm}[t]
\vspace{-3ex}
\caption{FeReX Feasibility Detection Algorithm}

\label{alg:feasibilitydetection}
\begin{mdframed}[linecolor=black,linewidth=1pt]
\textbf{INPUT: } The $M \times N$ Distance Matrix $DM$ to be implemented by each cell, which includes $K$ FeFETs, with a current range $CR=[C_1,C_2,\cdots C_n]$ allowed to flow through each FeFET\\
\textbf{OUTPUT: } \textit{Feasible Region}  or $False$\\

\For{i \textbf{from} 0 \textbf{to} M-1}{
    \For{j \textbf{from} 0 \textbf{to} N-1 } {
        DMCurs[i,\ j] $\gets$ \textbf{DecomposeDM}(K,\ DM[i,\ j],\ CR)
    }
    Searchlines[i] $\gets$ \textbf{Backtracking}(DMCurs[i])
}

FeasibleRegion $\gets$ \textbf{AC3}(Searchlines) \\

\If{\textit{Feasible Region} not exist}{
\textbf{return} \textbf{False}
}
\textbf{return} \textit{Feasible Region}\\
\end{mdframed}

\end{algorithm}

The implemented CSP with above three constraints has many classical solution methods. Here, we choose Backtracking~\cite{bitner1975backtrack} and AC3~\cite{mackworth1977consistency,soto2013hybrid}
to determine whether a set feasible FeFET currents under all \sch and \sto conditions exists, 
as illustrated in Alg.~\ref{alg:feasibilitydetection}. 
The initial current set DMCurs is enumerated per the first constraint, and
Backtracking and AC3 are utilized to effectively address the second and the third constraints, respectively. 
If the objective is to obtain all possible current sets, AC3 can be replaced by backtracking. 
The output of the algorithm is the \textit{Feasible Region}, which consists of the filtered sets
$DMCurs$ satisfying the three constraints. 

\begin{figure}[!t]
  \centering
  \includegraphics[width=1\columnwidth]{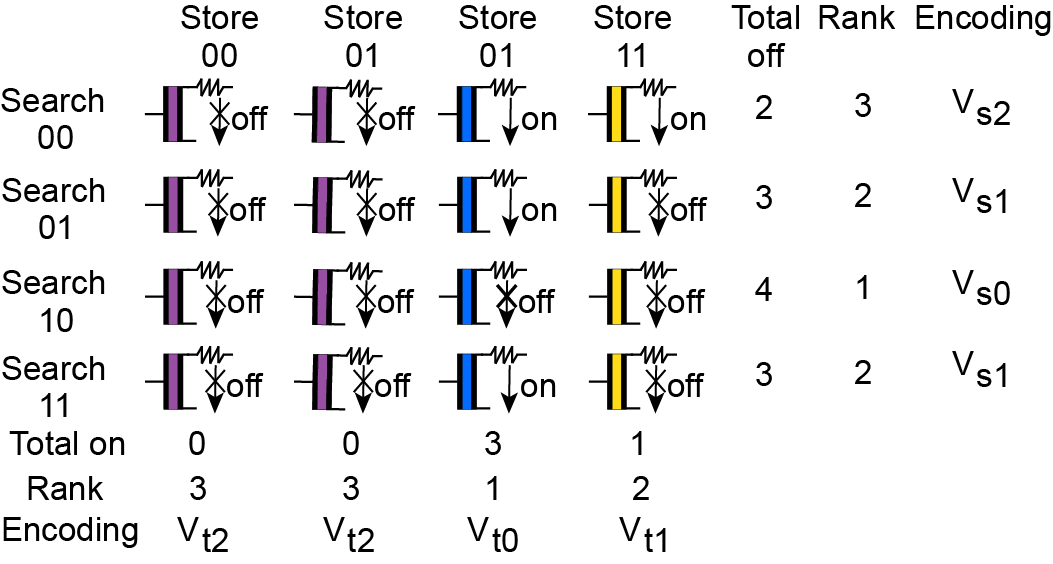}
  \caption{Encoding  \textit{Feasible Region} from  algorithm. \ref{alg:feasibilitydetection} to  the store/search voltage configurations for a single FeFET device.}
  \label{fig:EncodingPostProcessing}
  
  \vspace{-2ex}
\end{figure}
\begin{table*}[t]
    \centering
    
    \caption{3FeFET3R 2bit Hamming Distance encoding table}
    \label{tab:3FeFET}
    \resizebox{0.95\textwidth}{!}{
        \tiny
        \begin{tabular}{@{}c|ccc|ccc|ccc@{}}
            \toprule
            &
            \multicolumn{3}{c|}{\textbf{Store Encoding}} & \multicolumn{6}{c}{\textbf{Search Encoding}} \\ 
            \cmidrule(l){2-4} \cmidrule(l){5-10}
            &
            \textbf{$V_{th,FET1}$} & \textbf{$V_{th,FET2}$} & \textbf{$V_{th,FET3}$} & \textbf{$V_{g,FET1}$} & \textbf{$V_{g,FET2}$} & \textbf{$V_{g,FET3}$} & \textbf{$V_{ds,FET1}$} & \textbf{$V_{ds,FET2}$} & \textbf{$V_{ds,FET3}$} \\
            \midrule
            "00" &$Vt_2$ & $Vt_2$ & $Vt_0$ & $Vs_2$ & $Vs_2$ & $Vs_0$ & $V$ & $V$ & $V$ \\    
            "01" &$Vt_2$ & $Vt_0$ & $Vt_2$ & $Vs_1$ & $Vs_0$ & $Vs_2$ & $2V$ & $V$ & $V$ \\   
            "10" & $Vt_0$ & $Vt_2$ & $Vt_2$ & $Vs_0$ & $Vs_1$ & $Vs_2$ & $V$ & $2V$ & $V$ \\  
            "11" & $Vt_1$ & $Vt_1$ & $Vt_1$ & $Vs_1$ & $Vs_1$ & $Vs_1$ & $V$ & $V$ & $2V$ \\  
            \bottomrule
        \end{tabular}
    }
    \vspace{-3ex}
\end{table*}
Fig.~\ref{fig:EncodingPostProcessing} demonstrates the post processing of the output \textit{Feasible Region} to obtain all the possible search and stored voltage encoding configurations for a single FeFET. 
For a feasible set DMCurs, during the stored $\sto$ encoding process, the numbers of ON states in all $\sto$ columns are counted and sorted. 
The $\sto$ columns with  higher ranks correspond to lower $V_{th}$ voltages.
During the search  $\sch$ encoding process, similarly, the numbers of  OFF states  in all $\sch$ rows are counted and sorted. 
The $\sch$ rows with  higher ranks  correspond to lower $V_{search}$ voltages.
The $V_{ds}$ encoding  corresponds to  non-zero values in DMCurs.

Tab.~\ref{tab:3FeFET} summarizes the encoding results for 2-bit Hamming Distance with the proposed \design circuit. \design iteratively increases the number of FeFETs within a cell, and determines that a 3FeFET3R cell structure is the optimal solution for the DM of 2-bit Hamming Distance. 
The FeFET is ON only if  $V_{t_i}<V_{s_j}$, where $i<j, i, j\in \{0,1,2\}$. This encoding scheme has also been extended to  other distance functions such as multi-bit Manhattan and multi-bit Euclidean. We leverage encoding of multi-bit Manhattan and multi-bit Euclidean in  Sec.~\ref{subsec:bench} for benchmarking.

    
\section{Evaluation \& Benchmarking}
\label{sec:eval}
In this section, we evaluate the FeReX using Cadence Virtuoso in terms of accuracy, robustness, and power consumption.
The Preisach FeFET model \cite{ni2018circuit} was adopted for FeFETs, while the 45nm PTM model \cite{vattikonda2006modeling} was used for all MOSFETs. Wiring parasitics for the 45nm technology node were extracted from DESTINY \cite{poremba2015destiny}. The operational amplifier (op-amp)  was based on the design from \cite{kassiri2013slew} and  scaled down to  45nm technology. 
\subsection{Array  Evaluation}
\begin{figure}[!t]
  \centering
  \includegraphics[width=\columnwidth]{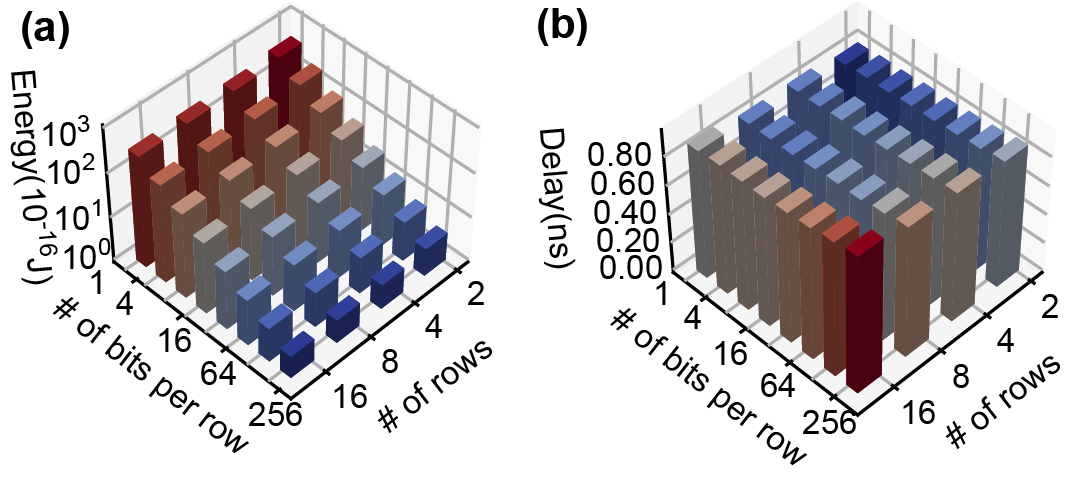}
  \caption{Search energy and delay of FeReX: (a) Energy  per bit and (b) delay with varying number of rows and dimensions.}
  \label{fig:EnergyAndDelay}
  \vspace{-2ex}
\end{figure}

Fig.~\ref{fig:EnergyAndDelay}(a) demonstrates that increasing the number of rows in the \design can reduce the average energy consumption per bit, since the power consumption of LTA  grows insignificantly as the number of rows increases.
The search delay consists of two  parts. 
About 60\% of the total delay comes from  ScL voltage stabilization associated with the op-amp, which is constrained by the op-amp's slew rate. 
The remaining delay associates with the LTA circuitry. As shown in Fig.~\ref{fig:EnergyAndDelay}(b), the total delay increases gradually as the FeReX array scales.

\begin{figure}[!t]
  \centering
  \includegraphics[width=\columnwidth]{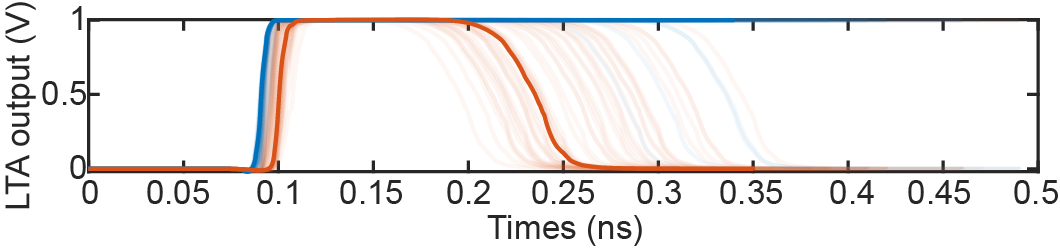}
  \caption{Monte Carlo simulations considering device-to-device variations: FeReX achieves 90\% accuracy in the worst search case
  of KNN workloads. }
  \label{fig:MC}
  \vspace{-2ex}
\end{figure}

To further validate the effectiveness of the proposed \design, 
we  conduct Monte Carlo (MC) simulation in the context of KNN, by taking device-to-device variation into account. 
Then, 
we benchmark \design with the vector-symbolic architecture (VSA) framework \cite{kleyko2022vector}, also known as the hyperdimensional computing (HDC).
Fig.~\ref{fig:MC} illustrates the MC simulations of FeReX with 100 runs. The device-to-device variation for the FeFET threshold voltage was set to 54mV \cite{soliman2020ultra}, and the resistance variation for the 1FeFET1R structure was extracted from fabricated data \cite{saito2021analog}, set to 8\%. The \design array level results demonstrate 90\% search accuracy when comparing the stored vectors with Hamming distances 5 and 6 to the query, representing the most challenging search cases\footnote{The closest Hamming distance between query and stored vectors is 5.}  of KNN when executing MNIST.
This performance results in only a 0.6\% accuracy degradation compared to the software-based implementation.

\begin{figure*}
    \centering
    \includegraphics[width=\linewidth]{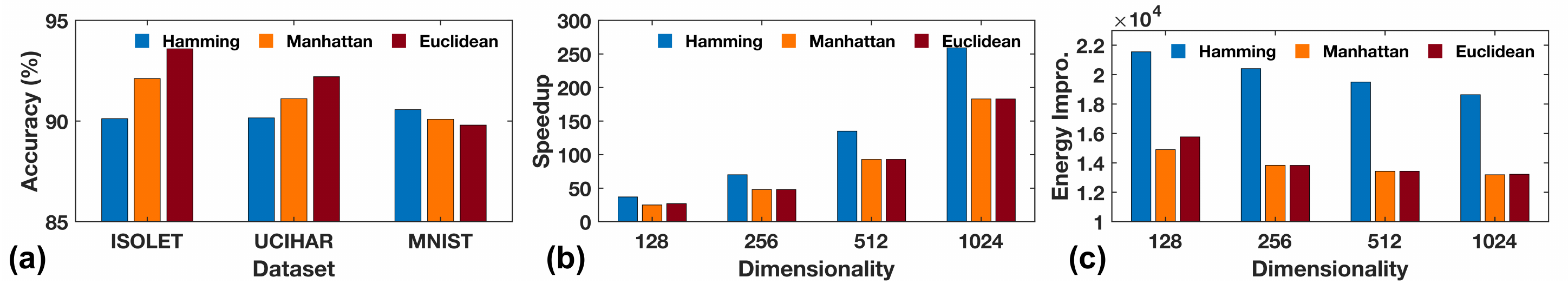}
    \vspace{-4ex}
    \caption{(a) Classification accuracy with different FeReX distance metrics. (b) Computation speedup and (c) energy efficiency improvement over  GPU implementation.}
    \vspace{-3ex}
    \label{fig:benchmark}
\end{figure*}

\subsection{Application Benchmarking}
\label{subsec:bench}
\textbf{

\begin{table}[t]
\centering
\caption{Datasets ($n$: feature size, $K$: number of classes)}
\label{tab:benchmark}
\resizebox{1\columnwidth}{!}{
\begin{tabular}{|c|ccccc|}
\toprule
\textbf{Dataset}& $n$ & $K$ & \shortstack{\textbf{Train}\\ \textbf{Size}} & \shortstack{\textbf{Test}\\ \textbf{Size}} & \textbf{Description}             \\ \midrule
\textbf{ISOLET} & 617               & 26         & 6,238               & 1,559              & Voice Recognition~\cite{Isolet}              \\
\textbf{UCIHAR} & 561               & 12          & 6,213                & 1,554               & Physical Activity Monitoring~\cite{anguita2012human} \\ 
\textbf{MNIST} & 784             & 10        &  60,000   & 10,000       & Handwritten Recongition~\cite{lecun1998gradient}\\ 
\bottomrule
\end{tabular}
}
\vspace{-2ex}
\end{table}}
we  briefly discuss the advantages of HDC benchmarking and its algorithmic flow. In HDC, low dimensional features are initially projected to high dimensional representations randomly, enabling \textit{holographicness} across the high dimensional feature vectors. HDC is pre-defined through a set of transparent operations, and due to its holographicness, it has been reported to be robust against hardware noise \cite{hernandez2021reghd}.

The algorithmic flow of HDC can  be categorized into three steps: first,  data is projected to  high dimension, as mentioned above. Second, single-pass training is performed, where the encoded high-dimensional vectors of a certain class are aggregated. Iterative training are conducted for higher algorithmic accuracy. Finally, during the inference phase of classification, the predicted class vector that has closest distance to the query vector is output using the configured FeReX distance function.

Here, we benchmark the proposed FeReX in the context of HDC with Nvidia 3090 GPU \cite{hernandez2021onlinehd} over three large-scale datasets given in Tab.~\ref{tab:benchmark}. By extracting the latency of the inference operations through \textit{Pytorch Profiler} package, the energy is obtained with the Nvidia System Management Interface. 
Fig.~\ref{fig:benchmark}(a) shows the accuracy of the reconfigurable search engine. Conventional  CIM-based HDC accelerator implements Hamming distance, yet different distance metrics may result in better accuracy across different datasets.
Fig.~\ref{fig:benchmark}(b) and (c) show the efficiency of the proposed  FeReX, showcasing up to 250x speedup and $10^4$ energy improvement over the GPU implementation.

\section{Conclusion}
\label{sec:conclusion}
In this paper, we propose \design, a FeFET-based AM for reconfigurable distance NN search. Based on  derived FeFET device and circuit constraints, \design filters and encodes feasible search and stored voltage configurations to implement a distance matrix of the target  distance function by addressing the constraint satisfaction problem. 
Evaluations at array level
validate the functionality and efficiency of the proposed  \design, and benchmarking results illustrate the improvement of  \design implementation over GPU.
To the best of our knowledge, this is the first   NVM based AM with reconfigurable search distance function, which will pave the way towards reconfigurable  AM designs for broader ranges of emerging applications.

\section*{Acknowledgements}
This work was supported in part by  National Natural Science Foundation of China (62104213, 92164203, 62122005), Zhejiang Provincial Natural Science Foundation (LD21F040003, LQ21F040006), National Key R\&D Program of China (2022YFB4400300), the Research Grant Council of HKSAR (27210321), the Croucher Foundation and the ACCESS—AI Chip Center for Emerging Smart Systems, sponsored by InnoHK funding, Hong Kong SAR. 
Liu was supported by CoCoSys, one of seven centers in JUMP2.0, a Semiconductor Research Corporation (SRC) program sponsored by DARPA. Imani was supported in part by National Science Foundation \#2127780, \#2319198,  \#2321840 and \#2312517, Office of Naval Research \#N00014-21-1-2225 and \#N00014-22-1-2067, the Air Force Office of Scientific Research \#FA9550-22-1-0253.


\bibliographystyle{IEEEtran}
\footnotesize{%
\bibliography{refs}
}
\end{document}